\documentclass[10pt,twocolumn]{article}

\usepackage[a4paper,margin=0.7in]{geometry}
\usepackage{graphicx}
\usepackage[style=numeric, backend=biber, natbib=true,
            maxcitenames=2, mincitenames=1,
            maxbibnames=5, minbibnames=5]{biblatex}
\usepackage{hyperref}
\usepackage{titlesec}
\usepackage{orcidlink}

\titleformat{\section}
  {\normalfont\large\bfseries}
  {\thesection}
  {1em}
  {}

\titleformat{\subsection}
  {\normalfont\normalsize\bfseries}
  {\thesubsection}
  {1em}
  {}

\usepackage[table]{xcolor}
\usepackage[most]{tcolorbox}
\usepackage{enumitem}

\newtcolorbox{rqsummary}[2][]{%
  enhanced, breakable,
  colback=gray!8,
  colframe=black!25,
  boxrule=0.4pt,
  arc=6pt,
  left=10pt, right=10pt, top=8pt, bottom=10pt,
  borderline={0.3pt}{0pt}{black!15},
  title={#2},
  title filled,
  titlerule=0pt,
  colbacktitle=black!85,
  coltitle=white,
  fonttitle=\bfseries\itshape,
  title after break={},
  #1
}

\def\circled#1{\raisebox{.5pt}{\textcircled{\raisebox{-.9pt} {#1}}}}

\title{\textbf{Evaluating Code Recommender Systems: A Review}}

\author{
    Daniel Borst\,\orcidlink{0009-0000-1622-4955},
    Stefan Sobernig\,\orcidlink{0009-0002-5018-7961} \\
    \textit{\small Institute for Complex Networks, WU Vienna,
    Welthandelsplatz 1, 1020 Vienna, Austria} \\
    \small\url{https://complex.wu.ac.at/}
}

\date{}

\begin{document}

\twocolumn[
\maketitle

\begin{center}
\begin{minipage}{\textwidth}
\begin{abstract}
\textbf{Context:} Code recommender systems (CRSs) are specialized
software systems operating on source artifacts to provide
automatically generated recommendations to software developers in all phases
of development. The goal is to improve software quality while enhancing
developers' efficiency, effectiveness, and
experience. \textbf{Problem:}~Despite the growing importance, little
is known about the state of evaluating these systems in a
human-centric manner.
\textbf{Research Approach:} We conducted a systematic literature
review to identify and to synthesize primary studies evaluating CRSs
(2017--2024). Ninety-two publications were included and subjected to
a systematic content analysis. \textbf{Results:} Our study confirms
that \textit{Offline Evaluations} are the most common,
system-centric evaluation type, whereas user-centric evaluations
(\textit{Online Evaluations}, \textit{User Studies}) are rarely
reported. Evaluations are concentrated on the \textit{Software Construction}
phase. Most studies explicitly report threats to validity, with
\textit{External} threats to study \textit{Materials} being the most
common. \textbf{Conclusion:} The state of evaluations on CRSs is
narrowly focused on single or a few system qualities. There is an
imbalance between a majority of system-centric evaluations and a
minority of user-centric evaluations. Combined, multi-type
evaluations are rarely reported.
\end{abstract}

\vspace{0.5em}

\end{minipage}
\end{center}

\vspace{1em}
]

\section{Introduction}
\begin{figure}[h!]
    \centering
    \includegraphics[width=1\linewidth]{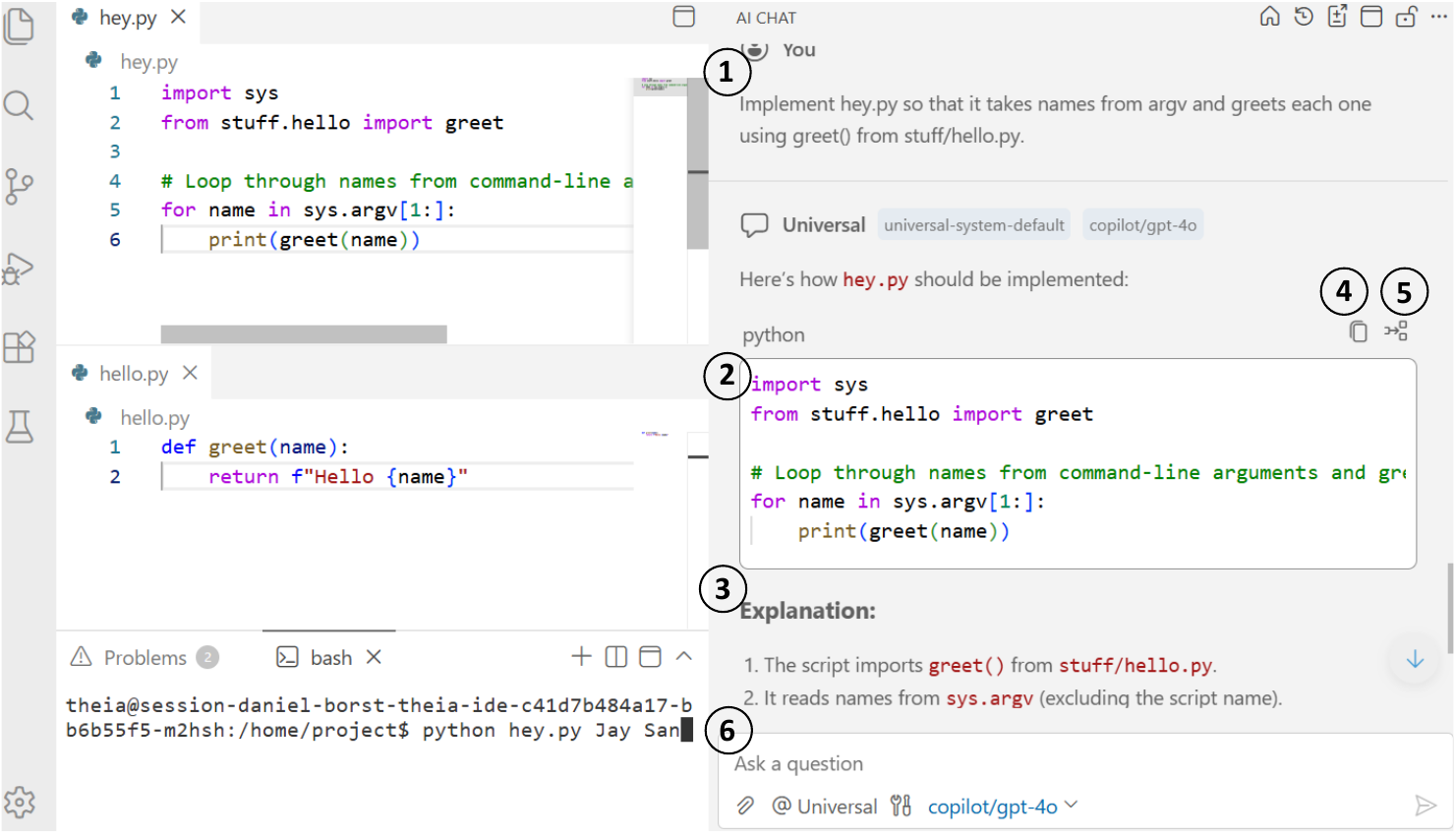}
    \caption{Example: The Eclipse Theia AI IDE~\cite{helming2025}
      offers a built-in code recommender system (here: the default
      Chat UI in front of an LLM-based chat agent driven by GPT-4o)
      that processes a natural-language prompt~\circled{1} to produce answers
      that include code blocks~\circled{2} as the code recommendation,
      surrounded by a natural language explanation~\circled{3}. The
      recommended code block can be copied~\circled{4}, inserted at the cursor
      location~\circled{5}, or refined through further prompting~\circled{6}.}
    \label{crs_ex}
\end{figure}

Code recommender systems (CRSs) are specialized software systems
designed to assist developers in various software engineering (SE)
activities~\citep{Robillard}. These systems provide automatically
generated recommendations as part of software language services in
software development systems, such as IDEs (see
Fig.~\ref{crs_ex}). Relevant language services as recommendation
channels encompass both syntactic aspects, like static program
analysis and semantic aspects, such as automated code
generation~\citep{Erdweg}. Depending on the supported activities, the
recommender methods and techniques used, the degree of automation, and
the interaction styles between software developers and CRSs, they are
referred to as programming assistants, code assistants, chatbots, or
code tools, among others. CRSs operate directly at the level of
artifact sources (source text, or diagram representations for models),
which means that their input and/or their output include, but are not
limited to, source fragments (e.g., the code block \circled{2} in
Fig.~\ref{crs_ex}).

Originating in the 1960s~\citep{Ossher}, CRSs have evolved into a
practical and accepted developer tooling. Large language models (LLMs)
have enabled systems that combine multiple recommendation approaches,
allowing for non-trivial operations such as code generation and fixing
source-code defects~\citep{She, Wang}. 

Despite the rapidly growing importance, little is known about the
state of research evaluating these systems and their quality in
use. Software engineering, including programming, is considered a
multidisciplinary and human-centric activity that involves complex social
interactions~\citep{Lenberg2015}. Comprehensive system evaluations,
therefore, involve a whole spectrum of system-centric and user-centric
evaluation targets in different lab and field contexts. \emph{So, how
  comprehensive is the reported evaluation practice that investiges
  software developers' observations and experiences of the promised and
  actual outcomes of applying code recommender systems?} Answering
this question is not only a prerequisite understanding the real and
potential impact of code recommender systems in
software-developing organisations. Mapping out and characterising this
evaluation practice helps explain the development,
proliferation, and persistence of influential,
decision-driving misconceptions about code recommender
systems~\citep{butler:2026:myths}. For example, do evaluations and the
evidence gathered reflect the fact that the major share of working
time and perceived workload of software developers is spent in
activities \emph{other than coding} during software construction (see
Myths 1 and 2 in~\citealt{butler:2026:myths})?

Certainly, relevant secondary, literature-based research
exists~\citep{Bauer,Almonte,Wang,Kotsiantis}. However, this prior work
has not put the spotlight on evaluating CRSs independent of their
underlying recommendation methods and techniques (including LLM and
non-LLM ones). The focus was rather on recommender
systems~\citep{Bauer} across various fields, in a more general
sense. Others have limited themselves to certainly related, but more
specific recommender systems in model-driven engineering
(MDE)~\citep{Almonte}, or have focused exclusively on CRSs based on
LLMs for code generation~\citep{Wang} and for assisted
programming~\citep{Kotsiantis}. On top, this prior research has not
been updated since 2023.

\emph{Therefore}, to provide the evidence missing between general RS
and specific CRSs, we report on a systematic literature review
(SLR; \citep{Kitchenham,Zhang,Wohlin}) on the state of evaluating code
recommender systems between 2017 and 2024. From an initial set of
4,126 publications, we applied predefined selection and quality
criteria, resulting in 92 primary studies subjected to a systematic
content analysis. We distill evidence on employed evaluation types
(\textit{Offline}, \textit{Online}, and \textit{User Studies}), the supported
software engineering areas, and recommender operations. Furthermore,
we analyze the employed metrics and synthesize reported threats to
validity associated with CRSs evaluations. To enable reproduction and
replication, all data, scripts, and materials are provided in a
publicly available research artifact~\citep{reppackage}.

In the remainder, Section~\ref{sec:bg} provides background on CRSs,
and their evaluation. Section~\ref{sec:rdesign} introduces the
research questions and documents the review
design. Section~\ref{sec:results} answers the research questions by
presenting the corresponding review results. Section~\ref{sec:threats}
outlines potential threats to the validity, followed by a discussion
of key insights in
Section~\ref{sec:discussion}. Section~\ref{sec:related_work} presents
related work and Section~\ref{sec:concl} concludes this review report.

\section{Background}\label{sec:bg}
Code recommender systems have been categorized along various
dimensions which can be broadly grouped into technical (e.g., model
architecture~\citep{Mavridou}, recommendation approach~\citep{Almonte}) and functional characteristics. This
work focuses on functional dimensions most relevant to the evaluation
of these systems, including classifications based on the supported
software processes~\citep{DelCarpio}, the supported working tasks of
developers~\citep{Leblanc, Wong, Hossami, Almonte, Mavridou}, and their operation mode~\citep{Leblanc, Almonte, Gasparic}. The supported software processes correspond to some of the SWEBOK~\citep{Bourque}including knowledge areas, including software (Sw.) requirements, Sw. design, Sw. construction, Sw. testing, and Sw. maintenance. Developers' working tasks have been considered at different levels of granularity. Generic task definitions differentiate between (code) complete, create, find, repair, and reuse tasks~\citep{Almonte, Mavridou}. Code recommenders operate in different modes from the
perspective of the interacting developer, including reactive, proactive, or mixed modes~\citep{Leblanc, Almonte, Gasparic}. In this literature study, CRSs aiming to support developers in all of the above processes, for all working tasks in all operational modes, are considered.

\subsection{Evaluating Code Recommenders}
\textbf{\emph{Evaluation}} is the systematic assessment of a code recommender's quality, which typically involves collecting data on phenomena external to the recommender, most importantly, from developers interacting with a recommender in the context of specific work processes. \emph{Validation} is defined as the systematic assessment of a recommender's quality through direct observation and measurement of phenomena internal to the recommender. Evaluation can include validation to ensure functional correctness. A recommender's quality in evaluation and in validation is determined in terms of one or multiple \emph{quality characteristics} (e.g., usability, maintainability; \citep{ISO25002:2024}). If not observable directly, a
quality characteristic is operationalized in terms of \emph{metrics}
(e.g., efficiency, satisfaction; \citep{NIST}). A metric is defined as
a subjective characterization of a quality characteristic based on
quantifiable and objective data. \emph{Measures} are measurement operations applied during data collection to determine the quantitative or qualitative value(s) of a metric (e.g., task completion time, number of errors;~\citep{Fenton, NIST}).

Evaluative research on recommender systems (RSs) distinguishes between
three types of evaluations: \textit{Offline Evaluations}, \textit{User Studies}, and \textit{Online Evaluations}~\citep{Shani,Zangerle,Freyne}.

\textbf{\emph{Offline evaluations}} are system-centric evaluations and
do not involve developers as system users or their interactions with
the recommender system~\citep{Bauer}. Emphasis is on evaluating the
underlying methods and techniques, such as algorithms and models, to
gain insights into the quality of recommendations. This can be
accomplished through computational evaluations conducted on historical
or synthetic data designed to simulate real-world
conditions~\citep{Shani, Almonte}. One example is the
HumanEval~\citep{MarkChen} benchmark, which comprises human-written
programming problems provided as natural-language prompts for
evaluating a recommender’s code generation
capabilities~\citep{Yan}. The results of offline evaluations are of a
quantitative nature~\citep{Shani, Bauer}, including
information-retrieval metrics such as precision, recall, and accuracy,
e.g., to quantify the match between generated recommendations and
reference solutions~\citep{Sun}.

Their cost-effectiveness and ease of execution render offline
evaluations the most prevalent evaluation type in RS
research~\citep{Shani, Bauer, Almonte, Proksch}. They are easy to
repeat, scalable, and allow for systematic adjustments of the
experimental setup, computational configurations, and runtime
settings. This enables extensive comparisons across multiple systems
under varied conditions and for fine-tuning them~\citep{Shani,
Bauer}. However, high-quality datasets are not always available or
can be challenging to access~\citep{Freyne}. Their lack of user
involvement means they cannot capture user-system interactions, which
restricts the type of recommender qualities which can be
investigated~\citep{Shani, Bauer}.

\textbf{\emph{User studies}} are user-centric evaluations and involve
human participants who perform predefined tasks that require interaction with a recommender system~\citep{Bauer}, e.g., locating
and fixing a defect in a code artifact~\citep{Campos}. The primary goal is to observe the user-system interactions under controlled conditions in laboratory or live settings. Through direct, real-time
observations, researchers can collect data on various aspects of the
RS~\citep{Shani}. This feedback includes quantitative data (e.g.,
task-completion time~\citep{Xu2}) and qualitative data via surveys,
questionnaires, or interviews~\citep{Bauer, Freyne}, e.g., on the perceived helpfulness~\citep{Campos} or the perceived usability of the
recommender~\citep{Ferdowsi}.

On the flip side, user studies can be time-consuming and
costly. Costs arise from setting up the evaluation environment (e.g.,
configuring software, installing sensors or logging tools) and from
recruiting and compensating participants~\citep{Bauer,
Proksch}. Additionally, user studies remain limited to a
small number of participants and tasks, limiting their scalability and
generalizability~\citep{Almonte, Proksch}.

\textbf{\emph{Online evaluations}} are user-centric evaluations and
collect feedback on a system's quality from users in the field
\citep{Bauer}. Unlike user studies, participants perform self-selected
tasks. User behavior and interactions with the system are logged to
collect quantitative data (e.g., recommendation acceptance rates~\citep{Sahoo}), complemented by qualitative insights
gathered through surveys, questionnaires, and interviews (e.g., willingness to use system~\citep{Kogler})~\citep{Bauer,
  Shani}. Online evaluations are capable of generating evidence of a
system's \emph{quality-in-use} under field conditions over an extended
period~\citep{Shani, Freyne}.

However, online evaluations can be resource-intensive due to
their complexity, requiring significant time and financial investment,
making them challenging to repeat. Deploying a system in a field
setting also presents practical challenges, including organizational
restrictions in the field and difficulties in recruiting
participants. Moreover, the field environment restricts data
collection compared to controlled settings~\citep{Almonte, Bauer}.

\subsection{Literature Reviews}\label{sec:bg:lr}
Literature reviews are secondary studies to gather and to synthesize
existing data and results on a specific topic from \textit{primary studies}. Literature reviews have become an established research
approach in SE~\citep{Kitchenham} to summarize existing evidence on the benefits and limitations of SE methods, techniques, and tools, to
identify gaps in current research, and to provide the motivation for
staging new research activities. Among others, two main types of
secondary studies in SE are discriminated: \textit{systematic mapping studies} (SMSs) and \textit{systematic literature reviews} (SLRs).

While both follow systematic principles and are supported by
established guidelines~\citep{Kitchenham,Wohlin,Zhang}, they serve
different purposes. SMSs aim to provide a broad overview of a research
area by classifying and clustering existing work. Their results can
reveal gaps in the literature and highlight areas where conducting an
SLR or new primary studies may be most valuable~\citep{Budgen}. In
contrast, SLRs aim to exhaustively identify all relevant primary
publications within a specific research area, including an assessment of their quality, with the objective of answering predefined RQs through systematic data extraction and
synthesis~\citep{Kitchenham}.
\begin{figure*}[htbp]
    \centering
    \includegraphics[width=\linewidth]{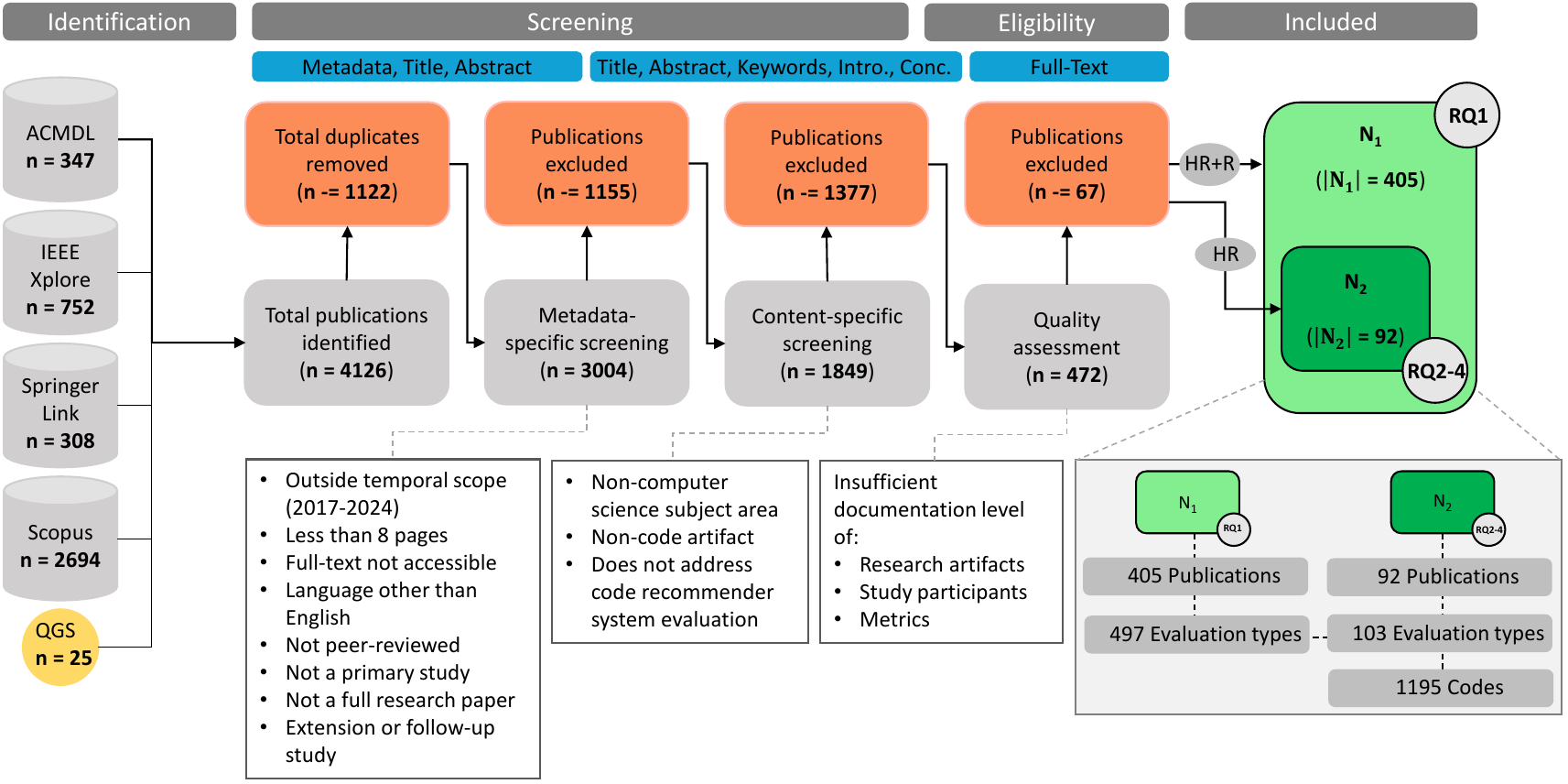}
    \caption{Stepwise process for selecting publications for review.}
    \label{SLR_over}
\end{figure*}
\section{Review Design}\label{sec:rdesign}

The goal of this study is to gain a deeper understanding of how code
recommender systems (CRSs) are evaluated. Prior work has broadly
reviewed the state and practice of evaluating recommender systems in
general (\citealt{Bauer}; see also Section~\ref{sec:related_work}) or
evaluating selected subtypes of code recommenders, such as code
generation using LLMs~\citep{Wang}, LLM-assisted
programming~\citep{Kotsiantis}, or model-driven software
development~\citep{Almonte}. Relevant findings from these studies
signal an overreliance on \textit{Offline Evaluations} compared to
\textit{User Studies} and \textit{Online Evaluations}. To provide a
comprehensive overview of CRS evaluation, we designed, planned, and
conducted an SLR to collect, classify, synthesize relevant data from
primary studies evaluating CRSs, and to analyze how different
evaluation types relate to one another. As an integral component of
the SLR approach, we performed a quality assessment, including a
review of the level of documentation, thereby providing insights into
the replicability of studies in this domain. The review design follows
the guidelines of~\citet{Kitchenham},~\citet{Zhang},
and~\citet{Wohlin}.

\subsection{Research Questions}

\begin{itemize}
    \item \textbf{RQ\textsubscript{1}} \emph{What is the prevalence of offline evaluations, user studies, and online evaluations for code recommender systems?}
    \item \textbf{RQ\textsubscript{2}} \emph{How are different evaluation types used to assess the quality characteristics of code recommender systems in software development?}
    \item \textbf{RQ\textsubscript{3}} \emph{What metrics are employed to assess the quality characteristics of code recommender systems?}
    \item \textbf{RQ\textsubscript{4}} \emph{Which threats to validity and limitations are
  reported in studies evaluating code recommender systems?} 
\end{itemize}

The RQs were addressed using two different publication sets ($N_1$,
$N_2$ in Fig.~\ref{SLR_over}, with $N_2 \subset
N_1$). RQ\textsubscript{1} investigates the prevalence of applied
evaluation types, which can be determined from study reports
addressing the evaluation of CRSs as either a minor and/ or major
contribution ($N_1$), that is, \textit{R(elevant)} and
\textit{H(ighly)R(elevant)} publications. Therefore, RQ1 was answered
using the full publication corpus ($|N_1] = 405$). In contrast,
RQ\textsubscript{2}--RQ\textsubscript{4} refer to specific evaluation
characteristics (e.g., metrics and threats to validity) that require
the evaluation to be reported in sufficient detail. These RQs were
therefore answered using the subset of HR publications addressing the
evaluation of CRSs as a major contribution ($|N_2| = 92$). This staged
qualification for the R and HR publication subsets was performed
during study selection. Two independent raters assigned relevance
labels, including conflict resolution. Refer to Sections~\ref{sec:rp}
and~\ref{sec:ms}, ``Thematic Relevance'' (C\textsubscript{10}), for
the details.

\subsection{Review Protocol}\label{sec:rp}
A review protocol established prior to executing the SLR is
available~\citep{slrprotocol}. For planing the SLR, relevant secondary
studies were identified using Google Scholar, with search terms
derived from our RQs~\citep{Wohlin}.
\paragraph{\textbf{Selection criteria}}
Of the ten selection criteria applied on retrieved publications, eight were metadata-specific (C\textsubscript{1}-C\textsubscript{8}) and two content-specific (C\textsubscript{9}-C\textsubscript{10}). Metadata considered were: C\textsubscript{1} publication date ($\geq 2017$)\footnote{The SLR covers publications from 2017 onward, as advancements during this period, particularly the integration of code embeddings and transformer-based models, improved the performance and capabilities of CRSs and increased their practical relevance and adoption in software engineering practices~\citep{Kotsiantis}.}, C\textsubscript{2} paper length ($\geq 8$ pages), C\textsubscript{3} full-text
accessibility, C\textsubscript{4} language (English), C\textsubscript{5} peer-reviewed, C\textsubscript{6} study type (primary), C\textsubscript{7}
publication type (full-research paper in conference, journal), C\textsubscript{8} originality (follow-up studies and extensions within the corpus were treated as a single publication). The two content-specific ones were:

\paragraph{Study context (C\textsubscript{9})} A publication must
explicitly address systems and operations that support software
engineering activities related to the process of creating or modifying
software language code. The assessment is based on the assignment to a
subject area category (e.g., Computer Science, Mathematics, Medicine)
and the artifact type investigated (Code, Data, Documentation). Only
publications assigned to the subject area Computer Science and that
investigate Code artifacts are included in the review.

\paragraph{Thematic relevance (C\textsubscript{10})}

A publication must make a qualified contribution to code recommender
systems and their evaluation. A code recommender system denotes any
software system that assists developers in software engineering
activities by providing automatically generated code
recommendations. We deliberately do not restrict the search based on
specific CRS characteristics (see Section~\ref{sec:bg}), such as the
underlying technology (e.g., LLMs) or interaction mode (e.g.,
proactive) \citep[Sec.~6.1]{slrprotocol}. This follows from our RQs and is in line with SLR
guidelines recommending the exhaustive identification of relevant
publications within a research area~\citep{Kitchenham}. We distinguish
between minor, major, and mixed contributions regarding design,
implementation, and evaluation of CRSs. The assessment is based on a
five-point ordinal relevance scale: A publication is \textit{N(ot)
  R(elevant)} (NR; does not address CRSs at all), \textit{S(lightly)
  R(elevant)} (SR; addresses designing and implementing CRSs as a
minor contribution, but there is no evaluation), \textit{M(oderately)
  R(elevant)} (MR; addresses designing, implementing, and evaluating
CRSs as minor contributions), \textit{R(elevant)} (R; designing and
implementing CRSs as a major contributions and their evaluation as a
minor one), and \textit{H(ighly) R(elevant)} (HR; addresses the evaluation
of CRSs as the major contribution). Only publications rated
\textit{R(elevant)} or \textit{H(ighly) R(elevant)} are included in
the review (see Fig.~\ref{SLR_over}).
The following examples illustrate the selection procedure for the three most critical relevance levels. Publications classified as \textit{MR}, for example, describe and evaluate techniques that can be used in CRSs but do not directly address a recommendation aspect as part of their contribution~\cite{Shi}. In contrast, \textit{R} publications focus on the development or improvement of a CRS and evaluate the resulting system, with the evaluation primarily serving to validate the proposed system~\cite{Dehaerne1, Chen}. Finally, \textit{HR} publications make the evaluation of CRSs their primary contribution, for example, by comparing multiple CRSs across various benchmarks~\cite{Zhang3} or by assessing the impact of CRSs when used in practice~\cite{Ferdowsi}.

\subsection{Planning the Search}
The first step in the process was to compile a corpus of reference
publications, referred to as \emph{quasi-gold standard}~\citep{Zhang},
used to iteratively define and continuously validate a search string for the main search. Reference publications were retrieved using a pilot search.

\paragraph{Pilot}
The pilot search involved a combination of automated search and
backward snowballing~\citep{Wohlin}. The automated search was conducted in Google Scholar with two search strings, developed using the PICOC method \citep{Petticrew}. To limit the scope, we considered only the first 30 pages of results and performed a preselection, resulting in 27 publications. Backward snowballing was performed using the secondary or literature studies identified during the preparation stage as the starting set. For the pilot study, we considered one iteration sufficient. This yielded 124 publications, which, together with the automated search results, resulted in 151 candidate publications. After removing 15 duplicates, metadata-specific criteria were applied by the first author as single reviewer, leading to the exclusion of 89 publications. The remaining 47 underwent content-specific screening by three independent reviewers. Ratings were compared and conflicts resolved in a pairwise
sessions between the reviewers. This screening led to the exclusion
of a further 22 publications.

\paragraph{Quasi-gold standard corpus and search engines}
25 publications formed the quasi-gold standard (QGS) corpus (see also
Fig.~\ref{SLR_over}). This paper collection consist of 18 conference
and seven journal articles that cover the entire temporal study
period, with at least two and up to five publications per year. These
works are published in 17 different venues by six distinct publishers: ACM (6 venues), IEEE (5), Elsevier (2), Springer (2), USENIX (1), and ACL (1). We identified four search engines for these publishers: ACM Digital Library, IEEE Xplore, Scopus, and SpringerLink. These search engines met the criteria outlined in the protocol, including full coverage of the time period, access to bibliographical metadata, and minimal content overlap.

\paragraph{Search string}
In line with \citet{Zhang}, we manually extracted candidate search
terms by analyzing the titles, abstracts, and keywords of all QGS
publications. Additionally, we incorporated terms derived from our
RQs. After cleaning the candidate terms, we categorized them to
construct an initial search string, which is then refined iteratively
through quasi-sensitivity analysis. The final search string comprised
27 distinct search terms , divided into two categories: \textit{evaluation} (4 terms) and \textit{tool \& action} (23) (see~\ref{asec: ss}).

All search engines supported the use of structured search
strings. The search was targeted to the title, abstract, and keywords of the publications. Additionally, initial filtering options provided by the search engines were applied, such as specifying the time scope (2017-2024) or restricting the publication language to full-text articles in English.

\subsection{Main Search}\label{sec:ms}
The main search was performed on November 4, 2024, and resulted in 4,101 hits
divided into four result sets: ACMDL (347), IEEE Xplore (752), Scopus
(2,694), and SpringerLink (308). In total, the search retrieved 24 of
the 25 QGS publications, yielding a quasi-sensitivity of 96\% (the one
 publication missing from the search results was included, nonetheless). Deduplication
identified a total of 1,122 duplicates (1,042 from the automated
process, 56 from the semi-automated process, and 24 QGS publications), leaving 3,004
publications for the actual selection phase (see Fig.~\ref{SLR_over}).

\paragraph{\textbf{Selection of primary studies}}
Eight metadata-specific criteria, which are based on objective
attributes, were evaluated on the 3,004 publications by the first
author, resulting in the exclusion of 1,155 publications. The two
content-specific criteria required subjective assessment and were
evaluated on each publication by two independent reviewers (see
C\textsubscript{9} and C\textsubscript{10} in Section~\ref{sec:rp}).

\textit{Study context (C\textsubscript{9})} was assessed by the two authors as
reviewers. The publication pool was divided into subsets, with each
subset initially screened by one reviewer. Those publications marked
for exclusion were considered for an additional review by the
second. Any disagreements that arose were resolved through a consensus
discussion. The two authors achieved an overall Percent
Agreement~\citep{Gwet} $p_a = 82.6\%$ (456 out of 522), which ultimately left 1,374 publications for further review.\footnote{The inter-rater reliability (IRR) measures used in this work were selected based on the measurement scale of the data (e.g., nominal, ordinal) and the rating procedure (e.g., two reviewers, three reviewers), following the guidelines by~\citet{Gwet}. Further details are provided in the review protocol~\citep{slrprotocol}.}

The evaluation of \textit{thematic relevance (C\textsubscript{10})}
was performed by five reviewers: the two authors and three
contributors (two PhD students, one from the same Institute as the authors and one from an
academia-industry project and a Master's student
experienced as a CS1 tutor). Each publication was assigned to two
independent reviewers using round-robin scheduling~\citep{Moseley} to
ensure balanced coverage across search engines and publication
years. After each evaluation round, the reviewer pairs compared the results on a given publication, and resolved any disagreements. Early rounds
included group meetings to establish a common understanding and
warming-up, whereas in later rounds consensus agreements were settled
directly between reviewers of overlapping assignments. The reviewers
achieved an overall agreement of Cohen’s Weighted Kappa~\citep{Gwet}
$\hat{\kappa_c} = 0.71$, ranging from 0.67 to 0.72 across reviewer pairs.

After having applied C\textsubscript{9} and C\textsubscript{10}, 1,377 publications (475 and 902, respectively) had been excluded and 472 were left for quality assessment.

\paragraph{\textbf{Quality assessment}}

The quality check aimed at the breadth and the depth of the
study-design documentation, including artifacts, participant details
(where applicable), and evaluation metrics. The quality assessment was performed by the two authors. The first author reviewed
the publication’s full text, as well as any supplementary materials,
and assigned ratings on a four-point ordinal scale ranging from \textit{Not Documented} to \textit{Fully Documented} for each dimension, based on predefined quality
criteria. Publications that fell below a defined quality threshold
were excluded. This threshold was set such that any publication receiving a rating below \textit{Documented} in the artifacts field (e.g., insufficient description of the study design, including objectives, variables, and procedures) or below \textit{Partly Documented} in either the participants (e.g., no information on study participants provided, if applicable) or evaluation metrics fields was excluded. Publications marked for exclusion underwent an
additional review by the second author. The two authors as reviewers achieved an overall Percent Agreement $p_a = 50.0\%$, with disagreements becoming resolved in
consensus-finding sessions. In total, 67 publications were excluded
leaving 405 publications included after quality assessment (see
Fig.~\ref{SLR_over}).

\begin{figure}[h!]
    \centering
    \includegraphics[width=1\linewidth]{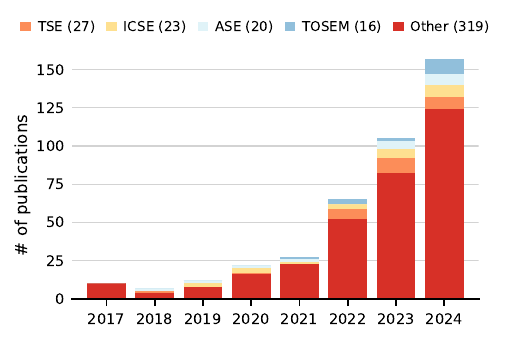}
    \caption{Number of included publications per year (in total:~405),
      along with the most prevalent publication venues.}
    \label{included_over}
\end{figure}

\paragraph{\textbf{Included publications}}
A total of 405 publications were included. Fig.~\ref{included_over} presents the
number of publications per year, along with the most prevalent
publication venues. Most of the publications (245) appeared in
conference proceedings, while 160 were published in journals. The
corpus spans publications from 132 distinct venues, with the top four
being TSE (27 publications), ICSE (23), ASE (20), and TOSEM (16) (see
Fig.~\ref{included_over}). The 319 \textit{Other} venues include JSS (13), FSE (13), ACL (13), IST (12), SANER (11), ICLR (11), ICPC (11), ISSTA (10), EMNLP (10), ICPS (9), EMSE (7), IEEE Access (7), PACMSE (6), NeurIPS (6), ASE Journal (6) MSR (6), and PMLR (5). The remainder of 111 venues contribute four or less publications to the corpus.

The publication timeline indicates a significant increase in publications starting in 2022, which continues to rise sharply until the end of the study period in 2024. This observation aligns with the findings of \citet{Kotsiantis}, who reported that CRSs are becoming increasingly mainstream and integrated into IDEs starting in 2021. Evaluative research followed. 

\subsection{Data Extraction and Content Analysis}\label{dataex}
Data extraction followed a hybrid coding approach that was primarily deductive, based on a predefined coding scheme, but allowed for inductive flexibility through an \textit{Other} category in which new themes could emerge~\citep{Fereday}. The scheme comprises six top-level
categories, with each top-level category representing a key
\emph{D}imension of evaluation studies
(D\textsubscript{1}--D\textsubscript{6}): Evaluation Type
(D\textsubscript{1}), Metrics and Measures (D\textsubscript{2}),
Software Engineering Area (D\textsubscript{3}), Code Recommender
System Operation (D\textsubscript{4}), Software Language Type
(D\textsubscript{5}), and Reported Threat Types
(D\textsubscript{6}). The full extraction form is provided in our
protocol~\citep{slrprotocol}. To align our findings to prior research, and allow
for systematic comparisons, we derived typologies from related studies.

\paragraph{\textbf{Categories and codes}} For D\textsubscript{1}, we adopted
the evaluation types by \citet{Bauer,Shani}: \textit{Offline
  Evaluations}, \textit{User Studies}, and \textit{Online Evaluations}
(see also Section~\ref{sec:bg}). A fourth evaluation type was adapted
based on the notion of a judgment study~\citep{Stol}:
\textit{Human-involved Offline Evaluation}. This type is defined as
\textit{Offline Evaluations} using datasets or benchmarks that were created using input by human participants, either for generating the dataset (human-CRSs interaction) or for judging results (e.g., correctness of a generated code snippet), that is, without live
user interaction during the experiment.

For metrics and measures (D\textsubscript{2}), we first considered
their source of documentation (D\textsubscript{2.1}): \textit{Defined by Authors}, \textit{From Other Publication}, \textit{Standard or Industry Benchmark}, and \textit{Not Indicated}. In addition, metrics were described by their measurement objective (D\textsubscript{2.2})~\citep{Wang,ISO25010:2023,ISO25019:2023}:
\textit{Acceptability}, \textit{Functional Suitability},
\textit{Performance Efficiency}, \textit{Reliability}, \textit{Safety
\& Security}, and \textit{Usability}. Finally, we considered the measurement object (D\textsubscript{2.3}) of
each metric and measure, grouped into one of: \textit{Recommendation},
\textit{Developer}, \textit{System}, and \textit{Work Process}.

Publications were also categorized by the software (Sw.) engineering areas
the CRSs are meant to operate in (D\textsubscript{3}):
\textit{Sw. Requirements}, \textit{Sw. Design},
\textit{Sw. Construct}, \textit{Sw. Testing},
\textit{Sw. Maintenance}, and \textit{Sw. Configuration Management}.
These were adapted from \citet{DelCarpio} and SWEBOK knowledge
areas \citep{Bourque} by replacing Sw. Process Learning
with Sw. Configuration Management.

To capture characteristic CRS operations (D\textsubscript{4}), we
adopted the typology by \citet{Almonte}, which distinguishes five
types: \textit{Complete}, \textit{Create}, \textit{Find},
\textit{Repair}, and \textit{Reuse}. For cataloging the software
languages and their artifacts subjected to the CRS operations
(D\textsubscript{5}), we distinguished between the language type
(\textit{Modeling} or \textit{Programming}), the application scope
(\textit{Domain-Specific} or \textit{Domain-Independent}), and the
notation type (\textit{Graphical}, \textit{Textual}, or
\textit{Tabular}).

Only threats to validity (D\textsubscript{6}) reported by the authors of the included publications were extracted and typed (D\textsubscript{6.1}) following the
scheme of \citet{Wyrich}: \textit{Internal}, \textit{External},
\textit{Construct}, and \textit{Conclusion} (see also
\citep{Jedlitschka}), and seven threat areas (D\textsubscript{6.2}) \citep{Siegmund}:
\textit{Participants}, \textit{Materials}, \textit{Task Design},
\textit{Operationalization and Measurement}, \textit{Research Design},
\textit{Procedure and Conduct}, and \textit{Data Analysis}.

\paragraph{\textbf{Coding procedure}} Coding involved the full texts
(incl. figures, tables, captions as well as appendices), while codes
where assigned to paragraphs or blocks of sentences as coding
units. The coded text segments were documented along with the assigned
codes in coding sheets (workbooks; see the research artifact~\citep{reppackage} as
well as by colored highlights in the full-text PDFs).

Data extraction and coding were performed by the first author on all
92 publications. The second author independently performed data
extraction on a specific subset of 25 out of 92 publications. The
subset size was determined based on the degree of rating deviations
between reviewers for selection criterion
C\textsubscript{10} (see Section \ref{sec:rp}). Ratings of thirteen publications showed a strong deviation ($\sim$ 15\%), i.e., those having received a ``Highly Relevant'' by one, and a second rating below
``Relevant'' by another reviewer. Transferring this share to the
remaining 79 publications, additional 12 publications (79 × 0.15) were
selected using stratified simple random sampling without
replacement~\cite{Cochran}. Three strata were defined (``QGS'', ``Highly
Relevant/Relevant'', ``Highly Relevant/Highly Relevant''), four
publications were randomly drawn from each.

An overall agreement of Kupper-Hafner $KH = 0.62$ was reached. Most disagreements occurred in category D2. We identified two main reasons. The first was the heterogeneity of documentation practices when extracting evaluation metrics. For instance, when a publication reported a composite metric (e.g., System Usability Score), publications varied in how they documented the composite and/or the individual items (e.g., individual questions). The second, the hybrid coding scheme for D2, included overlapping or emerging subcategories which were applied differently by the authors during data extraction. During conflict resolution, these cases were discussed, a common understanding was reached, and the agreed changes were systematically applied to all subsequent coding decisions and incorporated as improvements to the protocol~\citep{slrprotocol}.

\section{Results}\label{sec:results}
\subsection{RQ1}
\begin{figure}[h!]
        \centering
        \includegraphics[width=\linewidth]{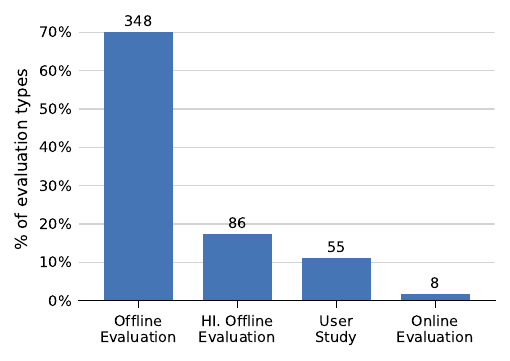}
        \caption{Prevalence of reported evaluation types across all included publications.}
        \label{prev_eval_meth}
\end{figure}

\textbf{What is the prevalence of offline evaluations, user
  studies, and online evaluations for code recommender systems?}
We extracted evaluation types reported in 405
publications, each reporting
at least one evaluation type. In total, 497
evaluation types were
identified, all of which could be assigned to one of the predefined
evaluation types (see Fig.~\ref{prev_eval_meth}).

\textit{Offline Evaluations} were the most frequently applied type,
accounting for 70.0\%. \textit{Human-involved Offline Evaluations}
were the second most frequent type (17.3\%), followed by \textit{User
Studies} (11.1\%). \textit{Online Evaluations} were rare (1.6\%).

\begin{figure}[h!]
        \centering
        \includegraphics[width=\linewidth]{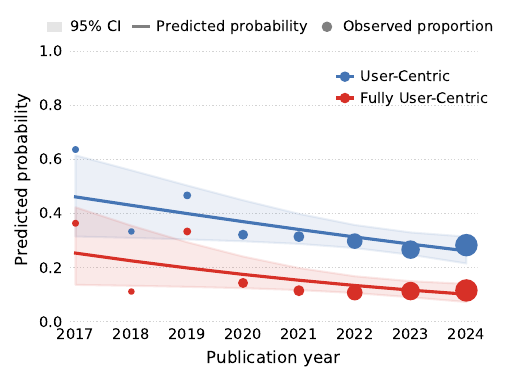}
        \caption{Predicted probability of a study being user-centric or fully user-centric by publication year, derived from logistic regression. Shaded areas represent 95\% confidence intervals. Dot size is proportional to the number of observations per year.}
        \label{log_reg_eval_meth}
\end{figure}

To analyze the temporal trend in employed evaluation types across the publication corpus, logistic regression was applied with publication year as the sole continuous predictor to two binary outcome variables (see Fig.~\ref{log_reg_eval_meth}): (1) user-centric evaluation (\textit{Human-Involved Offline Evaluation}, \textit{User Study}, or \textit{Online Evaluation}) versus system-centric evaluation (\textit{Offline Evaluation}), and (2) fully user-centric evaluation (\textit{User Study} or \textit{Online Evaluation}) versus offline-based evaluation (\textit{Offline Evaluation} or \textit{Human-Involved Offline Evaluation}). The odds ratio (OR) quantifies the multiplicative change in the odds of the outcome per one-unit increase in publication year \citep{Hailpern2003}. For both outcome variables the logistic regression results are statistically significant (p = 0.023, p = 0.026). User-centric evaluations show an OR of 0.882 (95\% CI: 0.791–0.983), indicating that the odds of a study employing a user-centric evaluation type decrease by 11.8\% per year. Similar results are obtained for fully user-centric evaluations, with a slightly lower OR of 0.854 (95\% CI: 0.743–0.981), corresponding to a decrease of 14.6\% per year.

A total of 94 publications reported two evaluation types, while no
publication reported more than two. The most frequent combination was
\textit{Offline Evaluation} together with \textit{Human-involved
  Offline Evaluation} (72 cases, 76.6\%). The second most common
combination was \textit{Offline Evaluation} with \textit{User Study}
(16 cases, 16.8\%). Less frequent were combinations of \textit{Offline Evaluation} with \textit{Online Evaluation} (3 cases, 3.2\%) and \textit{Human-involved Offline Evaluation} with \textit{User Study} (3 cases, 3.2\%).

Across the corpus, 24 publications were identified as members of a publication family (e.g., extensions, follow-ups), each explicitly referencing prior work. Within these families, 75\% of subsequent publications employed the same evaluation type(s) as their predecessor. In 25\% of subsequent publications, an additional or new evaluation type was introduced (e.g., progressing from \textit{Offline Evaluation}~\citep{Nguyen19} to \textit{Offline Evaluation} combined with \textit{User Study}~\citep{Phuong22}). Across all publication families, the number of distinct evaluation types did not exceed two.

\subsection{RQ2}
\textbf{How are different evaluation types used to assess the quality characteristics of code recommender systems in software development?}  In total, we
assigned 125 SE area codes, with each evaluation type in
a publication contributing at most one entry per SE area.

\begin{figure}[h!]
    \centering
    \begin{minipage}{0.45\textwidth}
        \centering
        \includegraphics[width=\linewidth]{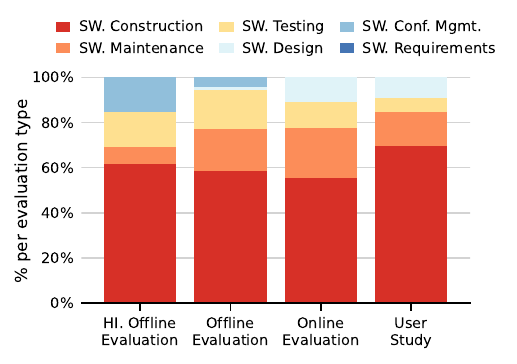}
        \caption{Software engineering areas across evaluation types at the method level (row-normalized percentages).}
        \label{cross_area_meth}
    \end{minipage}\hfill
    \begin{minipage}{0.45\textwidth}
        \centering
        \includegraphics[width=\linewidth]{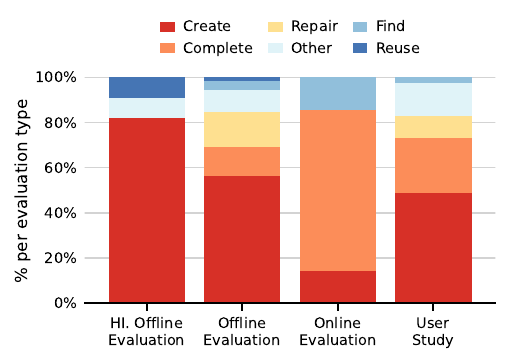}
        \caption{CRS operations across evaluation types at the method level (row-normalized percentages).}
        \label{cross_op_meth}
    \end{minipage}
\end{figure}

\textit{Sw. Construction} was the most
prevalent SE area (61.6\%), followed by \textit{Sw. Maintenance}
(16.8\%) and \textit{Sw. Testing} (13.6\%). \textit{Sw. Design}
(4.0\%) and \textit{Sw. Configuration Management} (4.0\%) were the
least frequent coded, while \textit{Sw. Requirements} was not coded at
all. Fig. \ref{cross_area_meth} shows the distribution of SE areas
across each evaluation type. Only \textit{Offline Evaluations} covered all
identified SE areas, although \textit{Sw. Design} was coded only once
(1.4\%). \textit{Online Evaluations} and \textit{User Studies} did not
address \textit{Sw. Configuration Management}, while
\textit{Human-involved Offline Evaluations} include no codes for
\textit{Sw. Design}. All evaluation types were primarily associated
with \textit{Sw. Construction} (55.6\%–69.7\%).

As for CRS operations, 130 codes were applied. The most frequent CRS operation was \textit{Create} (53.8\%),
referring to systems that help generate new artifacts from scratch,
followed by \textit{Complete} (18.5\%), which help complete existing
source artifacts, and \textit{Repair} (11.5\%), which help fixing defects in
existing source artifacts. CRS operations \textit{Find} (3.8\%) and
\textit{Reuse} (1.5\%), which support artifact discovery and
integration, appeared rarely. Fig.~\ref{cross_op_meth} displays the
operation frequencies split across the evaluation
types. \textit{Offline Evaluations} covered all predefined CRS
operations, whereas \textit{User Studies} did not include
\textit{Reuse}. \textit{Online Evaluations} did not address
\textit{Repair} or \textit{Reuse}, and \textit{Human-involved Offline
  Evaluations} lacked \textit{Repair} and
\textit{Find}. \textit{Create} was most frequently observed in all
evaluation types except \textit{Online Evaluations}, where
\textit{Complete} was most common.

Sixteen evaluation-type occurrences were coded as \textit{Other},
reflecting six distinct operations added to the coding scheme: Analyze
(5), Diagnose (4), Modify (3), Translate (2), Optimize (1).

94.2\% of evaluations focused on \textit{Programming} languages (e.g.,
Java, Python). In contrast, \textit{Modeling} languages (e.g., UML)
represented only 5.8\%. \textit{Domain-Independent} languages (e.g.,
UML, Python) were examined in 90.4\% of the evaluation types, while
\textit{Domain-Specific} languages (e.g., Ansible) appeared in only
9.6\%. Regarding notation, \textit{Textual} representations (e.g.,
Java) were the most common, occurring in 96.2\% of the evaluation
types, followed by \textit{Graphical} notation (e.g., UML) at 3.8\%.

\subsection{RQ3}
\textbf{What metrics are employed to assess the quality characteristics of code recommender systems?}
In total, the review yielded 393 unique metrics in 556 occurrences. 53
metrics were found reported in more than one publication.
From the unique metrics identified, 62.9\% were \textit{Defined by Authors}, 27.2\% originated \textit{From Another Publication} (third-party),
7.7\% came from other sources coded as \textit{Other} (software tooling, e.g., static analyzer) and 2.2\% are metrics from \textit{Standards or Industry Benchmarks} (e.g., NASA-TLX~\citep{Hart}, AttrakDiff~\citep{Hassenzahl}).

The most prevalent objective was \textit{Functional Suitability} (50.7\%), measuring the degree to which a measurement object (MO) meets its stated or implied requirements.
Followed by \textit{Usability} (17.4\%), which measures the extent to which the MO is understandable, learnable, and usable to achieve goals efficiently and satisfactorily. The newly defined objective \textit{Maintainability} (10.8\%) ranked third, while \textit{Acceptability} (6.3\%), \textit{Safety \& Security} (4.5\%), and \textit{Performance Efficiency} (4.1\%) were less common. Rarely reported objectives included, \textit{Cognitive Capabilities} (2.3\%), \textit{System Interaction} (2.2\%), \textit{Reliability} (1.4\%). In total, 471 (84.7\%) matched one of the six predefined measurement objectives. Under hybrid coding,  97 (16.8\%) were initially coded as \textit{Other}. In a next step, we introduced four new objectives to qualify the \textit{Other} matches:  \textit{Maintainability} (60), \textit{Cognitive Capabilities} (13), \textit{System Interaction} (12).

Each metric occurrence was also associated with an object under
measurement. Nearly half of the metrics, 244 in total, addressed the
object \textit{Recommendation} (43.9\%), that is, the measurement
targeted the respective code recommendation itself. The remainder
of metrics distributed across \textit{Developer} (33.3\%),
\textit{System} (12.6\%), \textit{Code Artifact} (7.2\%), and the
\textit{Work Process} (3.1\%). 516 metrics were mapped to predefined
codes, while 40 were initially coded as \textit{Other}, then captured under
the new label \textit{Code Artifact} (a code recommendation adapted by a developer and subsequently evaluated).

From the 393 unique metrics, 13 were reportedly used in the context of
more than four distinct primary studies: BLEU (15), pass@k (11), precision (10), recall (9), task completion time (9), accuracy (8), mean reciprocal
rank (MRR) (8), acceptance rate (6), correctness (6),
ROUGE (6), METEOR (6), CodeBLEU (5), and perceived readability
(5).

These most frequently used metrics have their origin in information
retrieval and natural-language processing (NLP) and are most commonly employed in \textit{Offline Evaluations}. They are genuinely
associated with the measurement objective \textit{Functional Suitability} and the measurement objects \textit{Recommendation} (e.g., BLEU, which measures similarity (n-gram overlap) between generated and reference text/code~\citep{Papineni}) and \textit{System} (e.g., pass@k, which captures the probability that at least one of the top-k generated code-candidates is correct (passes all ground-truth test cases)~\citep{MarkChen}).

The second most frequent measurement object, \textit{Developer}, focuses on the impact of the CRS on the end user. This impact can be measured either directly, using metrics like task completion time (\textit{Usability})~~\citep{Xu2} and acceptance rate (\textit{Acceptability})~\citep{Sahoo}, or indirectly through numerous, mostly non-standard subjective metrics capturing
perceptions: In \textit{Human-involved Offline Evaluations}, \textit{Developers} judge
aspects of the CRSs, such as a generated code snippet, without direct
interaction. In contrast, in \textit{User Studies} or \textit{Online Evaluations},
\textit{Developers} usually report their experience during or right after
having interacted with the CRSs. This type of metric covers various
measurement objectives, such as \textit{Functional Suitability} (e.g.,
perceived code correctness~\citep{Nguyen2}), \textit{Usability} (e.g., perceived ease of system use~\citep{Weber}), and \textit{Maintainability} (e.g.,
perceived code readability~\citep{Yuan}). Perceptional metrics are also often expressed through standardized composite scores, such as SUS~\citep{Brooke1996} or AttrakDiff score~\citep{Hassenzahl}; or ad hoc derivatives therefrom.

\begin{figure}[h!]
    \centering
    \includegraphics[width=1\linewidth]{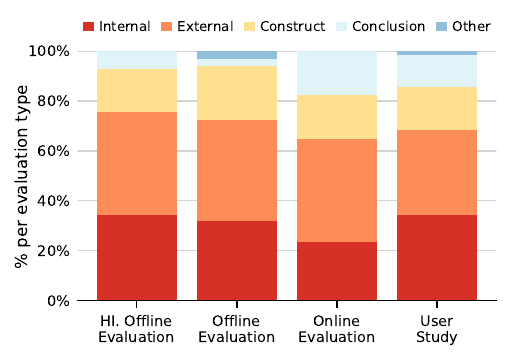}
    \caption{Threat types across evaluation types at the method level (row-normalized percentages).}
    \label{cross_type_meth}
\end{figure}

\subsection{RQ4}

\textbf{Which threats to validity and limitations are reported in studies evaluating code recommender systems?}
Our review and content analysis revealed threats to validity reported by the authors in 79 of the 92 publications (85.9\%), whereas no author-reported threats were
identified in the remaining 13 publications (14.1\%). Publications without reported threats occur at comparable rates across all four evaluation types (10-14\%; stratified rates). Of these 13, eleven were published in conference proceedings, two in journals.

Across the corpus, we assigned a total of 361 codes for threat type
and threat area. In decreasing order, the threat types
\textit{External} (38.8\%), \textit{Internal} (32.4\%),
\textit{Construct} (19.4\%), and \textit{Conclusion} (7.2\%) were
coded. Among the threat-reporting publications, \textit{External}
appeared in 94.9\%, while \textit{Internal} threats were reported
in 74.7\%. Initial coding of label \textit{Other} occurred eight times
in eight publications. In terms of hybrid coding, these were reviewed
to become grouped into two new sub-categories: \textit{Reliability} (four codes across two publications) and \textit{Replicability} (four codes across four publications). All reported \textit{Replicability} threats stem from studies evaluating LLM-based CRSs~\citep{Tran2024, Mousavi2024}, addressing the difficulty of replicating results due to the non-deterministic nature of LLMs. Commonly reported threats include \textit{External} threats related to study \textit{Materials}, such as the employed tools, software, or datasets (e.g., limited generalizability of datasets or benchmarks restricted to specific programming languages or CRS functionalities~\citep{Fakhoury2024, Du2024}) and \textit{Construct} threats related to \textit{Operationalization and Measurement}, concerning the definition and measurement of variables (e.g., adequacy of the measures or metrics used to operationalize a quality characteristic~\citep{Zhang3, Leclair2020}). 

Fig.~\ref{cross_type_meth} presents the distribution of threat types
across evaluation types. On average, \textit{User Studies} reported
4.9~$\pm$~2.2 (mean ± SD) threats to validity, followed by
\textit{Human-involved Offline Evaluations} with 4.1~$\pm$~1.8, \textit{Offline
Evaluations} with 3.7~$\pm$~1.4, and \textit{Online Evaluations} with
3.4~$\pm$~1.1. \textit{External} threats were reported at comparable
rates across methods (34.2\%–41.5\%), as were \textit{Construct}
threats (17.1\%–21.5\%). In contrast, \textit{Internal} and
\textit{Conclusion} threats showed greater
variation. \textit{Internal} threats were reported with similar
frequency in \textit{User Studies} (34.2\%), \textit{Human-involved Offline Evaluations}
(34.2\%), and \textit{Offline Evaluations} (31.7\%), but less often in \textit{Online
Evaluations} (23.5\%). \textit{Conclusion} threats were most prevalent
in \textit{Online Evaluations} (17.7\%), followed by \textit{User Studies} (12.1\%) and
\textit{Human-involved Offline Evaluations} (7.2\%), and were rarely reported
in \textit{Offline Evaluations} (2.7\%).

\begin{rqsummary}{Key results}
\begin{enumerate}[label=(\textbf{\arabic*}), wide, labelwidth=!, labelindent=0pt]
\item \textit{Offline Evaluations} are the most widely reported evaluation type (70\% of the reviewed primary studies).
\item Less than 1 in 3 adopt evaluation types with some human participation
  (i.e., \textit{Human-involved Offline Evaluations, User Studies,
    Online Evaluations}).
\item Code recommendation systems under evaluation focus on supporting
  the phase of \emph{Software Construction} (62\%) and \emph{Create}
  operations (54\%).
\item Frequently adopted metrics are information-retrieval ones (e.g.,
precision, recall, accuracy) and NLP ones (e.g., BLEU, ROUGE), to
operationalize the top-3 measurement objectives: \textit{Functional
  Suitability}, \textit{Usability}, and \textit{Maintainability}.
\item Threats to validity were reported in 86\% of publications. 3 in
  5 report \textit{External} and/or \textit{Internal}, only 1 in 4
  documents threats to \textit{Construct} and/or \textit{Conclusion}
  validity.
\end{enumerate}
\end{rqsummary}

\section{Threats to Validity}\label{sec:threats}

\paragraph{Formulation of research questions} This SLR represents the
first secondary study focusing on the evaluation of code recommender
systems (CRSs) independent from the underlying recommendation methods and techniques, for various types of source artifacts. The formulation of the four research questions (RQs) may be biased by subjective judgment and prior knowledge of the authors. To mitigate this threat, we consulted closely related secondary literature~\citep{Bauer,Wyrich}. The RQs were then developed using the
PICOC model \citep{Petticrew} based on insights reported from these
(\emph{construct validity}).

\paragraph{Search process} No search strategy can guarantee exhaustive
coverage of relevant primary studies~\citep{Kitchenham}. Limitations
may arise from incomplete keyword representation in the search
string. To reduce this threat, we employed a comprehensive strategy
based on the QGS approach \citep{Zhang}, combining automated searches,
backward snowballing, and a pilot study. In addition, our review was
restricted to peer-reviewed venues only. While excluding relevant
non-peer-reviewed and gray literature, it also ensures a certain
baseline quality of the included publications. The search was
performed in November 2024. Considering the reported publication
trend, any update to studies newer than this may revise the findings
(\emph{external validity}).

\paragraph{Study selection} Selection bias represents another
threat. The selection procedure combined automated and manual
processes. Automated processes may falsely exclude publications due to
incomplete metadata, whereas manual processes can introduce subjective
judgment and researcher bias. To address this, automated exclusion was
applied only when certainty was given, while cases such as duplicate
removal without DOIs were manually verified. Manual screening followed
a predefined protocol, with additional reviewers involved in decision
validation, inter-rater reliability (IRR) checks, and consensus
finding. An independent assessment of any selection bias is possible
by our publicly available research artifact~\citep{reppackage}
(\emph{internal validity}).

RQ\textsubscript{1} was answered using the full set of
publications classified in selection criterion C\textsubscript{10} as \textit{R(elevant)} or \textit{H(ighly) R(elevant)} ($|N_1| = 405$), whereas  RQ\textsubscript{2}--RQ\textsubscript{4} require evaluation characteristics to be reported in sufficient detail and where consequently answered using the subset of \textit{HR} publications (($|N_2| = 92$)) (see Section~\ref{sec:rdesign}). The classification underlying both sets is based on reviewers' judgments of the thematic relevance of a publication and is therefore affected by rater subjectivity. Consequently, publications could be misclassified in two ways: (1) excluded from $N_1$ even though R or HR (2) excluded from $N_2$ when in fact HR. Either case would bias the findings for the corresponding RQs by omitting eligible publications. To mitigate this risk, relevance labels were assigned independently by two reviewers, yielding a reviewer agreement of ${\kappa} = 0.75$ for $N_1$ and ${\kappa} = 0.53$ for $N_2$. Labels were assigned by four pairs of raters, requiring consistency within and across pairs to be established throughout the rating process. Publications with mismatched ratings were re-examined, and disagreements were resolved through consensus. In the early rounds, consensus was reached through discussions involving all reviewers, while in later rounds, disagreements were resolved within the rater pairs (\emph{construct validity}).  

\paragraph{Data extraction and analysis} Manual interpretation during
data extraction and analysis may introduce errors due to
misinterpretation, omission, or ambiguity in the primary studies. This
is particularly relevant as many extracted variables are nominal
without intrinsic ordering (e.g., measurement objective, CRS
operation). To mitigate this threat, we developed predefined
classification schemes informed by frameworks such as SWEBOK
\citep{Bourque} and prior secondary studies (\emph{construct
  validity}). Extraction was performed by a single reviewer, with 27\%
of the data independently double-coded. For the remaining
publications, extraction issues were discussed collaboratively among
the reviewers to reach consensus (\emph{internal validity}). The temporal trend analysis based on logistic regression relies on the assumption of independence of observations across publication years. This assumption may not hold, as for example studies from the same research group or publication venue may exhibit correlated evaluation practices. To mitigate this, extension or follow-up studies within the publication corpus were identified and treated as a single publication during data extraction (construct validity). Furthermore, the narrow time range (2017–2024) and the heavy concentration of observations in 2022–2024 limit the generalizability of the observed trends (external validity).

\section{Discussion}\label{sec:discussion}
The objective of this systematic literature review (SLR) was to
provide a comprehensive overview of the current state of evaluative
research on code recommender systems (CRSs). Insights from the field
include the types of evaluations used, the technical and
contextual scope of code recommendations, the metrics applied, and the commonly addressed threats to validity.

\paragraph{\textbf{Offline evaluations on the rise}}
The number of publications increased substantially during the observed period, from 10 in 2017 to 157 in 2024. Growth was modest until 2021, before rising sharply from 2022 onward. Despite this overall growth, the temporal trend analysis reveals that the odds of studies employing user-centric and fully user-centric evaluation types declined significantly over time (11.8\% and 14.6\% per year, respectively), suggesting that the increase in publications is mainly driven by system-centric evaluations. \textit{What factors might explain this trend?} We outline three possible explanations. First, \textit{Offline Evaluations} are the most cost-effective and easiest to execute among the evaluation types~\citep{Bauer, Almonte}, based on a large number of established benchmarks such as HumanEval~\citep{MarkChen} that are accessible through platforms such as Hugging Face and enable extensive cross-system comparison via leaderboards. Second, the already higher barrier of user-centric evaluations~\citep{Bauer, Proksch} has been further raised by the introduction of LLMs into CRSs, as their nondeterminism requires additional study design considerations to ensure reliable results~\citep{Baltes2026}. Third, as further illustrated in the \textit{Solitary Evaluations} paragraph below, the literature recommends combining all evaluation types, progressing from \textit{Offline Evaluations} towards \textit{User Studies} and \textit{Online Evaluations}~\citep{Almonte, Bauer, Shani}. As our corpus covers only the early years of LLM-based CRSs, the current prevalence of \textit{Offline Evaluations} may reflect the early research stages. More user-centric evaluations might be expected as the field matures. These explanations are not mutually exclusive and may jointly contribute to the observed trend.

\paragraph{\textbf{Offline with humans}}
\textit{Offline Evaluations} are the most widely reported evaluation
type, accounting for 70\% of the reviewed primary studies. This
observation is consistent with findings from the broader recommender
systems domain (68\%;~\citep{Bauer}) and from the subfield of
recommender systems for model-driven engineering
(60\%;~\citep{Almonte}). 

\textit{Offline Evaluations}, however, come in a hybrid type:
\textit{Human-Involved Offline Evaluations} represent the second most
prevalent evaluation type in our review (17\%). In this evaluation
type, \textit{Developers} judge aspects of the CRSs, such as a
generated code snippet, without direct interaction. This involvement
enables researchers to generate human-centric insights alongside
system-centric evaluation, with comparatively less effort and
resources required compared to \textit{User Studies} and
\textit{Online Evaluations}. In particular, three-quarters of publications
combining two evaluation types supplemented an \textit{Offline
  Evaluation} with a \textit{Human-involved Offline Evaluation},
suggesting that this evaluation type is gaining importance in CRS
research. \textit{User Studies} and \textit{Online Evaluations} remain clearly outnumbered: Only 1 in 7 primary studies falls into one of the two groups.

As detailed in the \textit{Offline evaluations on the rise} paragraph, the odds of studies employing user-centric evaluation types declined significantly over time. This indicates that the existing dominance of system-centric evaluations has further increased over the observed period (2017–2024), and suggests that this trend may persist if current practices continue.

Even though \textit{Human-Involved Offline Evaluations} provide valuable human-centered insights, they do not capture the behaviors of
real-world actors (both user and systems) in a real-world environment. The steeper decline in fully user-centric evaluations (i.e., \textit{User Studies} and \textit{Online Evaluations}) is therefore particularly concerning, as these are the only evaluation types that directly inform our understanding of developer practices and the real-world quality of CRSs.

\begin{rqsummary}{Implications}
\begin{itemize}[wide, labelwidth=!, labelindent=0pt]
\item Researchers evaluating a CRS should not limit themselves to system-centric evaluations (i.e., \textit{Offline Evaluations}). Consider user-centric ones (i.e., \textit{Human-Involved Offline Evaluations}, \textit{User Studies}, \textit{Online Evaluations}) also at an early research stage.
\item \textit{Human-Involved Offline Evaluations} serve as a way to generate human-centric insights alongside system-centric evaluation, with comparatively less effort and resources required.
\end{itemize}
\end{rqsummary}

\paragraph{\textbf{Solitary evaluations}}
Approximately 23\% (94) of the reviewed publications combined evaluation types, representing a higher proportion than for evaluative research on recommenders in general (14\%;~\citep{Bauer}). Approximately 6\% of the reviewed publications belong to a publication family, yet 75\% of subsequent publications within these families employed the same evaluation type(s) as their predecessor. Evaluations involving combined types, whether within a single publication or across a publication family, are limited to a maximum of two. Researchers on CRSs do not apply family of evaluations,
starting from \textit{Offline}, then going to \textit{User Studies},
and concluding with \textit{Online Evaluations}. This is in sharp
contrast to the recommendations on combining the three evaluation
types~\citep{Almonte, Bauer, Shani}. For example, \textit{Offline Evaluations}
are typically regarded as the first step. They are used to filter out
non-performing approaches, leaving a smaller pool of candidates for
further investigation. \textit{User Studies} are then used to explore specific
aspects of an RS or to compare systems against each other or to
establish an initial baseline. \textit{Online Evaluations} are conducted
towards the end of an evaluation phase. The goal is to compare the
complete system in the field over a prolonged period.

In summary, our review documents that CRS evaluations remain too
narrowly focused, demonstrating a clear imbalance between the large
number of system-centric evaluations and the limited number of
user-centric ones, consistent with findings from related fields
\citep{Bauer, Mavridou, Almonte}.

\begin{rqsummary}{Implications}
\begin{itemize}[wide, labelwidth=!, labelindent=0pt]
\item Researchers should employ the full spectrum of evaluation types when evaluating a CRS, progressing from \textit{Offline Evaluations} to \textit{User Studies} and concluding with \textit{Online Evaluations}.
\item \textit{Human-Involved Offline Evaluations} can serve as an intermediate step between \textit{Offline Evaluations} and \textit{User Studies}, providing initial human-centric insights that can inform the design of subsequent \textit{User Studies}.
\item Results across evaluation types may diverge or complement each other and should be interpreted together to obtain a more comprehensive understanding of the quality and impact of a CRS.
\end{itemize}
\end{rqsummary}

\paragraph{\textbf{Normal-angle versus  wide-angle lenses}}
Our analysis shows that among publications employing a single evaluation type, 89\% of system-centric evaluations address one or two measurement objectives (e.g., \textit{Functional Suitability}~\citep{Liu2, Zhang3, Chakraborty}), while 65\% of user-centric evaluations cover three or more (e.g., \textit{Functional Suitability}, \textit{Usability}, \textit{Maintainability}, \textit{Acceptability}~\citep{Weber, Alizadeh, Campos}). Publications combining user-centric and system-centric evaluations consistently address at least two measurement objectives, with 45\% covering three or more (e.g., \textit{Functional Suitability}, \textit{Usability}, \textit{Acceptability}~\citep{Abid}). Note that the number of measurement objectives covered does not in itself reflect an evaluation’s quality or completeness. However, the results suggest that user-centric and mixed evaluation designs tend to adopt a broader perspective, capturing multiple facets of CRSs.

\paragraph{\textbf{Out of the BLEU}}
We observed a relatively high prevalence of 17.4\% of primary studies (28\% of system-centric evaluations)
employing NLP metrics (e.g., token-matching ones such as BLEU, ROUGE,
METEOR). This is unexpected and surprising, because these artifact-based metrics have
been proven to be poor proxies for measuring quality characteristics
of CRSs~\citep{Evtikhiev, Mastropaolo, Zhou23, Zhuo24}. These metrics
rely on the assumption that a higher textual similarity between a
generated artifact and a reference artifact implies higher
quality~\citep{Mastropaolo}. However, they fail to capture semantic
equivalence, as they do not account for implementation diversity
(e.g., variable naming), and may therefore rank non-functionally
equivalent code higher than functionally equivalent
code~\citep{Zhou23,Mastropaolo}. In addition, their application
requires high-quality reference datasets, which can be difficult and
costly to obtain~\citep{Zhuo24, Mastropaolo}, and several studies
report a weak correlation with human judgment~\citep{Zhou23,
Evtikhiev}. Even within this class of metrics, there are substantial
inconsistencies. For example, up to six different BLEU variants
are reported, each with different scores for identical data. This not
only threatens construct validity but also limits comparability between
studies~\citep{TaoWei, ShiEnsheng, WeiTao}. 

\begin{rqsummary}{Implications}
\begin{itemize}[wide, labelwidth=!, labelindent=0pt]
\item Researchers should critically reflect on the use of token-matching metrics (e.g., BLEU, ROUGE, METEOR) for CRS evaluation and instead prioritize execution-based ones (e.g., pass@k \citep{Zhou23}), which directly measure code functionality and are robust to implementation diversity. 
\item The development of artifact-based metrics specifically designed for code-related tasks (e.g., CodeBLEU \citep{Ren}, CodeBERTScore \citep{Zhou23}) requires further research. Once validated, such metrics can serve as an alternative to execution-based metrics.
\item  Artifact-based metrics should be complemented by human-centric insights (e.g., \textit{Human-Involved Offline Evaluations}) to enable correct interpretation of results.  
\end{itemize}
\end{rqsummary}

\begin{figure}[htbp]
    \centering
    \includegraphics[width=1\linewidth]{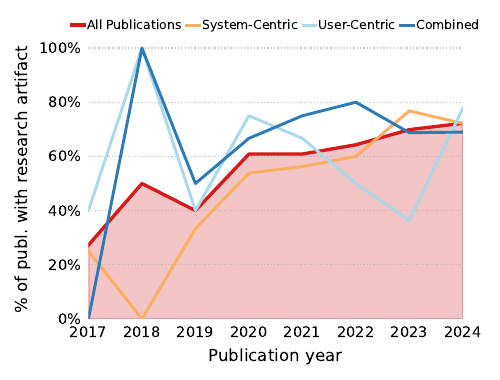}
    \caption{Share of publications providing research artifacts over time, grouped by evaluation type category.}
    \label{accessible_research_artifacts}
\end{figure}

\paragraph{\textbf{Replicability}}
For a quality assessment of the primary studies, we
reviewed both the publications themselves and any external
research artifacts available (i.e., supplementary artifacts to replicate a study
including material collections and publication supplements). Our
analysis revealed that 67\% of all publications (472)
provided an external research
artifact. Fig.~\ref{accessible_research_artifacts} illustrates the
share of publications providing research artifacts, categorized into
system-centric (i.e., \textit{Offline Evaluations}), user-centric
(i.e., \textit{Human-involved Offline Evaluations}, \textit{User
  Studies}, and \textit{Online Evaluations}), and combined evaluations
(i.e., publications including both system- and user-centric methods)
over time.

The results show a steady increase in the provision of
research artifacts, increasing from 27\% in 2017 to 72\% in 2024. This
trend is visible in all categories. Our findings align with those
of \citep{Liu}, who analyzed research artifacts in top software
engineering (SE) conferences (ASE, FSE, ICSE, and ISSTA) from 2017 to
2022 and reported an overall prevalence of 68\%, increasing from 60\%
to 81\% during that period. Another study \citep{Vega}, focusing on
human-oriented experiments in SE journals between 2000 and 2022, also
observed an increasing trend in artifact availability, particularly
after 2020, with a lower prevalence (40\%).
This difference is likely due to the inclusion of earlier publication
years of lower artifact availability. In addition, their focus on
human-oriented experiments may have further contributed to this
difference, as our data show that publications involving humans (i.e., user-centric evaluations) have a lower prevalence of available research artifacts (59\%) compared to system-centric (68\%) and combined (71\%)
evaluations.

\begin{rqsummary}{Implications}
\begin{itemize}[wide, labelwidth=!, labelindent=0pt]
\item To facilitate replication and comparison of CRS evaluations, and in line with the open science movement~\citep{Mendez2020}, researchers should provide for complete replication packages alongside their publications.
\item Replication packages should contain full experimental materials (e.g., datasets, tasks), infrastructure (e.g., benchmarks, telemetry software), and data (e.g., raw data, analysis scripts), except where ethical or privacy constraints apply (e.g., sensitive user data).
\item Providing complete replication packages, in particular for user-centric evaluations, can reduce the barriers (e.g., cost, time) of setting up such evaluations and thereby help address the growing imbalance between system- and user-centric evaluation in the CRS field.
\end{itemize}
\end{rqsummary}

\paragraph{\textbf{Validity reporting}}
Approximately 86\% (79/92) of the reviewed publications reported
threats to validity. These numbers match findings known from empirical
software engineering research, e.g., on code comprehension
experiments~\citep{Wyrich}. The absence of or lack of threats reports may be due to missing or ignored reporting guidelines, or limitations imposed by the outlets (page limits in conference articles; notably, 11 of the 13 publications without threat reports were conference papers). The absence of reported threats to validity is distributed across all four evaluation types at comparable rates (10-14\%), indicating this is not a limitation of any specific evaluation type. 

The distribution of threat types (i.e., \textit{External},
\textit{Internal}, \textit{Construct}, and \textit{Conclusion}
validity) aligns with the findings of \citep{Wyrich}. We also found
comparable shares of threats related to the areas of
\textit{Operationalization and Measurement}, \textit{Research
  Design/Procedure/Conduct}, \textit{Task Design}, and \textit{Data
  Analysis}. However, the primary studies on evaluating CRSs discussed
\textit{Material}-related threats more frequently and
\textit{Participant}-related threats less frequently than the studies
reviewed by \citep{Wyrich}. This can be explained by the most common
evaluation type in CRS research, \textit{Offline Evaluations} (70\%),
which do not involve human participants as a threat
area. \textit{Material}-related threats, being the most frequently
reported in our review, have the strongest impact on \textit{External}
validity. This can be explained by the strong dependence of
system-centric \textit{Offline Evaluations} on the models or datasets
used for evaluation. Furthermore, we identified a distinct set of \textit{Replicability} threats, unique to publications evaluating LLM-based CRSs, stemming from the non-deterministic behaviour of the underlying model.  

Additionally, we confirm the inconsistent use of threat-to-validity
categories similar to those identified by \citep{Wyrich}. We observed and
report several cases of misclassification that challenge our
analysis. Examples include confusing \textit{Construct} validity and
\textit{Internal} validity, as well as \textit{Construct} validity and
\textit{External} validity (see~\citep{slrprotocol} for the details).

\begin{rqsummary}{Implications}
\begin{itemize}[wide, labelwidth=!, labelindent=0pt]
\item Researchers should report threats to validity regardless of outlet type or page constraints.
\item Researchers evaluating LLM-based CRSs should follow established guidelines \citep{Baltes2026} to make claims auditable and results easier to replicate and compare.
\item Researchers should provide references or explanations for the used classification of threats to validity to prevent confusion and misclassification.
\end{itemize}
\end{rqsummary}

\section{Related Work} \label{sec:related_work}
Relevant systematic literature reviews (SLRs) on general recommenders
and specific subtypes of code recommenders are reported in
\citep{Bauer,Hou,DelCarpio}. All three reviews adopted the guidelines
by \citet{Kitchenham}. \citet{Bauer} retrieved 339 publications and
ultimately included 57 studies published between 2017 and 2022 to
investigate the state of evaluation research on recommender
systems. They listed the datasets used in these studies, classified
the metrics according to specific recommendation tasks (e.g., ranking,
rating prediction, relevance), and identified common metric
combinations. Additionally, they provided an overview of the types of
experimental designs employed. In their SLR, \citet{Hou} analyze the
role of large language models (LLMs) in software engineering
(LLM4SE). Using an automated search in combination with snowballing,
they identified 232,330 publications (2020--2024), of which 395 were ultimately
included in the review. Their analysis focused on the types of LLMs,
the employed datasets, fine-tuning techniques, evaluation metrics, the software engineering activities, and underlying tasks supported by
LLMs. Metrics were categorized according to the problem types
addressed (e.g., recommendation, classification), and key metrics were
identified for each task category. \citet{DelCarpio} examined the
field of SE assistants for broader, non-code-related working
tasks. From a search corpus of 4,246 publications, 40 publications
published between 2013 and 2023 were included for review. Their review
investigated implementation technologies, assistant types, interaction
types, supported software processes, evaluation types, and the metrics
employed.

Four related systematic mapping studies (SMSs) are reported
in~\citep{Almonte, Wang, Wyrich, Mavridou}. \citet{Almonte} focus on
CRSs in the model-driven engineering (MDE) domain. The SMS is designed
and conducted in line with the guidelines from~\citet{Petersen} and~\citet{Wohlin2}. The start set comprised 1,456 publications
(2004--2020), 66 publications were selected. The authors structured
the analysis along four dimensions: domain, tooling, recommendation,
and evaluation.  \citet{Wang} conducted a review of 20 publications
(2018--2023). Their SMS focuses on code generation using CRSs and is
driven by LLMs. Their analysis covers both the application areas of
LLM-based code generation and the evaluation of the generated
code. \citet{Mavridou} provide an overview of recommender systems for
programmers. Following the PRISMA-ScR guidelines
\citep{prisma_scoping}, the authors identified 745 and reviewed 50
publications (2015--2024). They included only publications that
explicitly addressed in-place code recommendations, excluding those
that focused on code generation or code search. CRSs are classified
according to the type of assistance, user input, output, and
methods. In addition, datasets and evaluation metrics are
cataloged. More recently, \citet{Wyrich} investigated
code-comprehension experiments and documented evidence on experiment
design, conduct, and reporting. They performed two iterations of
backward and forward snowballing~\citep{Wohlin}, identifying 95
publications (1979--2019). Our work complements these previous studies by including all CRSs regardless of their underlying recommendation method (e.g., code generation) or technique (e.g., LLM-based), and by focusing exclusively on their evaluation.

Research artifacts are essential to replication including
updates~\citep{Huotala}. Among the related studies, research artifacts
are only provided by \citep{Wyrich,Bauer}. There are also three
secondary studies outside of the SLR and SMS spectrum~\citep{She, Odeh, Dehaerne} that were consulted to identify primary studies for the pilot search (see Section~\ref{sec:rdesign}).

\section{Conclusion}\label{sec:concl}
We conducted a systematic literature review (SLR) synthesizing 92
primary studies (2017–-2024) evaluating code recommender systems. The
results indicate that system-centric \textit{Offline Evaluations} are
the most commonly reported evaluation type, whereas user-centric
approaches (i.e., \textit{Human-involved Offline Evaluations},
\textit{User Studies}, and \textit{Online Evaluations}) remain
underrepresented. Furthermore, most evaluations focus on the
\textit{Software Construction} phase and supporting developers in
\textit{Create} operations. Researchers predominantly employ
information retrieval and NLP-based metrics to operationalize the most
common measurement objectives: \textit{Functional Suitability},
\textit{Usability}, and \textit{Maintainability}. The majority of
studies report threats to validity, with \textit{Internal} and
\textit{External} threats being the most prevalent, followed by
\textit{Construct} and \textit{Conclusion} validity threats.

The dataset underlying this review is publicly
available~\citep{reppackage}, enabling further analyses of aspects such
as dataset characteristics, recruitment strategies, ethical
considerations, and the application of established best practices and
guidelines.

Further, our analysis demonstrates that the majority of new publications employ \textit{Offline Evaluations} while the proportion of user-centric evaluations has declined over time. We outlined three possible explanations for this trend, but further investigation is needed to confirm the underlying causes and to identify strategies to reverse it. One strategy could be the establishment of a methodical evaluation framework. Unlike the broader recommender systems domain, the CRS field currently lacks a widely agreed-upon framework to guide researchers in evaluation design decisions~\citep{Hou, Nascimento}. Such a framework could not only promote more standardized and comparable evaluation practices, but also encourage researchers to employ the full spectrum of evaluation types and to examine how results from different types may diverge or complement one another. In future work, based on a continued, in-depth content analysis of the reviewed publications and their research artifacts, we will extract reusable decisions on designing evaluations as a first building block for such a framework.

\section*{CRediT authorship contribution statement}
\textbf{Daniel Borst:} Conceptualization,  Investigation, Data curation, Formal analysis, Visualization, Writing - original draft, Writing - review \& editing.
\textbf{Stefan Sobernig:} Conceptualization, Data curation, Writing - original draft, Writing - review \& editing, Supervision, Funding acquisition.

\section*{Declaration of competing interest}
The authors declare that they have no known competing financial interests or personal relationships that could have appeared
to influence the work reported in this paper.

\section*{Data availability}
The data that support the findings of this study are openly available in Zenodo at \url{https://doi.org/10.5281/zenodo.19605843}.

\section*{Acknowledgements}
This work was supported by the Austrian Research Funding Association (FFG) within the funding programme KDT (call 2022) [grant number FO999902654].

\appendix
\section{Search String}
\label{asec: ss}

The full search string applied across all databases:

\begin{quote}
\small
(\texttt{"evaluat*"} OR \texttt{"assess*"} OR \texttt{"study"} OR \texttt{"experiment"}) \\
\hspace*{1em} AND \\
\hspace*{1em} (\texttt{"github copilot"} OR 
\texttt{"ai pair programmer"} OR \texttt{"code assistant"} OR 
\texttt{"openai codex"} OR \texttt{"query recommendation"} OR 
\texttt{"code completion"} OR \texttt{"code intelligence"} OR 
\texttt{"code recommendation"} OR \texttt{"refactoring recommendation"} 
OR \texttt{"code generation"} OR \texttt{"import generation"} OR 
\texttt{"commit message generation"} OR \texttt{"code comment generation"} 
OR \texttt{"code documentation generation"} OR 
\texttt{"automatic program synthesis"} OR 
\texttt{"assert statement recommendation"} OR \texttt{"code summarization"} 
OR \texttt{"code summarisation"} OR \texttt{"language workbench"} OR 
\texttt{"bug fixing"} OR \texttt{"type recommendation"} OR 
\texttt{"modelling chatbot"} OR \texttt{"code synthesis"})
\end{quote}

\printbibliography

\end{document}